\documentclass[12pt]{revtex4}
\usepackage{dcolumn}
\usepackage{graphicx}
\usepackage{amsmath}
\usepackage{amsfonts}
\usepackage{amssymb}
\usepackage{psfrag}
\usepackage{wrapfig}
\usepackage{subfigure}
\usepackage{makeidx}
\usepackage{bm}
\usepackage{epsf}
\usepackage{hyperref}

\begin{document}

\title{Static spherically symmetric star in Gauss-Bonnet gravity}

\author{ Kang Zhou$^{1}$, Zhan-Ying Yang$^{1}$\footnote{Email:zyyang@nwu.edu.cn},
De-Cheng Zou$^{1}$ and Rui-Hong Yue$^{2}$\footnote{ Email:yueruihong@nbu.edu.cn}}

\affiliation{ $^{1}$Department of Physics, Northwest University, Xi'an, 710069, China\\
$^{2}$Department of Physics, Northwest University, Xi'an, 710069, China}
\date{\today}

\begin{abstract}
\indent

We explore static spherically symmetric stars in the Gauss-Bonnet gravity without
cosmological constant, and present an exact internal solution which attaches to the exterior vacuum solution
outside stars.
It turns out that the presence of the Gauss-Bonnet term with a positive coupling constant
completely changes thermal and gravitational energies, and the upper bound of red shift of
spectral lines from the surface of stars. Unlike in general relativity, the upper bound of
red shift is dependent on the density of stars in our case. Moreover, we have proven that
two theorems for judging the stability of equilibrium of stars in general relativity
can be hold in Gauss-Bonnet gravity.
\end{abstract}



\maketitle

\section{Introduction\label{intro}}
\indent
Up to now, many quantum theories of gravity have been proposed. The
most promising candidate among them is the superstring/M-theory. In string
theory, extra dimensions were promoted from an interesting curiosity
to a theoretical necessity since superstring theory requires an
eleven-dimensional space-time to be consistent from the quantum
point of view \cite{Horava:1996ma,Lukas:1998qs,Randall:1999ee,Randall:1999vf}. It caused a
renewed interest about the general relativity in more than 4
dimensions. On the other hand, in recent
years another renewed interest has grown in higher order gravity,
which involves higher derivative curvature terms. Among the higher curvature
gravities, the most extensively studied theory is so-called Gauss-Bonnet
gravity \cite{Lanczos:1938sf,Lovelock:1971yv},
which can be naturally emerged when we want to generalize
Einstein’s theory in higher dimension by keeping all
characteristics of usual general relativity excepting the linear
dependence of Riemann tensor.
In addition, Gauss-Bonnet term occurs in the effective low-energy action of
superstring theory \cite{Lust:1989tj,Alekseev:1997wy,Maeda:2009uy,Elizalde:2007pi}.
Hence, the Gauss-Bonnet gravity provides a promising framework to study curvature
corrections to the Einstein-Hilbert action. In Gauss-Bonnet gravity,
the exact static and spherically symmetric black hole solutions were
found in \cite{Boulware:1985wk,Wheeler:1985qd,Wheeler:1985nh},
their thermodynamics have been investigated \cite{Cai:2001dz,Cai:2003gr,Cvetic:2001bk,Zou:2010tv}
and the slowly rotating black hole solutions
have been obtained in \cite{Kim:2007iw,Zou:2010dx,Zou:2011wd,Yue:2011zz}.

As well known, in general relativity, there is an exact solution
of gravitational fields with a spherically symmetric static star,
which takes the form of the exterior Schwarzschild solution of
vacuum outside the star \cite{Wald:1984rg}.
It implies that the exterior Schwarzschild solution can be naturally
emerged by a suitable distribution of matter. Thus, we are convinced
of the existence of the Schwarzschild black hole, since it is formed
from the gravitational collapse of a heavy star.
An interesting question will be, what's the solution of gravitational
fields with a spherically symmetric
static star in higher dimensional Gauss-Bonnet gravity,
which relates to the solution of black
holes, if matter can disperse to the extra dimension?

The objective of this paper is to find a solution and discuss the
physical properties of static spherically symmetric stars in Gauss-Bonnet
gravity without cosmological constant. The dimension of space-time is
assumed as $D\geq5$ since Gauss-Bonnet term yields
nontrivial dynamics. We try to investigate the spherically symmetric system
consisting of static perfect fluid, and give an exact solution.
In order to find the effect of Gauss-Bonnet term, we discuss thermal and gravitational
energies, the distribution of pressure of stars, and the upper bound of red shift of spectral
lines from the surface of them. Meanwhile,
we also study that whether two theorems for judging the stability of equilibrium in general
relativity can be hold in $D\geq5$ Gauss-Bonnet gravity.

This paper is organized as follows. In Section 2, we give the
equations for stars' structure in $D\geq5$ Gauss-Bonnet gravity without
the cosmological constant, and find a solution. In
Section 3, we investigate the effects of Gauss-Bonnet term. The
stability of equilibrium of stars is demonstrated in Section 4.
Some remarks are given in Section 5.

\section{Differential equations and solutions for star structure}  
\label{22s}
The action of Gauss-Bonnet gravity in $D$-dimensional space-time can be written in the form of
\begin{eqnarray}
S=-\frac{1}{2\kappa}\int{{d^D}x\sqrt{-g}(R+{\alpha}{\cal L}_2)+{S_{matter}}},\label{eq:1a}
\end{eqnarray}
where the coupling constant  $\alpha$
can be regarded as the inverse string
tension and we assume $\alpha>0$ in this paper.
Here the second order (Gauss-Bonnet) term of Lagrangian is given by
\begin{eqnarray}
{\cal L}_2 =R_{abcd}R^{abcd}-4R_{ab}R^{ab}+{R^2}.\nonumber
\end{eqnarray}
Varying the action Eq.~(\ref{eq:1a}), we obtain the field equations as
\begin{eqnarray}
{G_{ab}} = G_{ab}^{( 1 )} + {\alpha}G_{ab}^{( 2 )}=
\kappa{T_{ab}},\label{eq:2a}
\end{eqnarray}
where
\begin{eqnarray}
G_{ab}^{(1)}&=&{R_{ab}} - \frac{1}{2}R{g_{ab}}, \nonumber\\
G_{ab}^{(2)}&=&2( - {R_{acde}}R_{~~~b}^{dec} - 2{R_{acbd}}{R^{cd}} -
2{R_{ac}}R_{~b}^{c} + R{R_{ab}}) - \frac{1}{2}{\cal L}_2{g_{ab}}.\nonumber
\end{eqnarray}
For future simplicity, we take coefficient $\alpha =\frac{\tilde{\alpha}
}{(D-3)(D-4)}$.

The metric for a static spherically symmetric system will be taken in the form of
\begin{eqnarray}
ds^2=- B(r)dt^2 + A(r)dr^2 + r^2d\Omega_{D-2}^2,\label{eq:4a}
\end{eqnarray}
where $A$ and $B$ are functions of $r$, and $d\Omega
_{D-2}^2$ represents the line element of a $(D-2)$-dimensional unit sphere
with $\Omega_{D-2}=2\pi^{(D-1)/2}/\Gamma[(D-1)/2]$.
The energy-momentum tensor is assumed to be that for a perfect fluid
\begin{eqnarray}
T_{\mu\nu}=(\rho+p)U_\mu U_\nu+pg_{\mu\nu},\label{eq:3a}
\end{eqnarray}
with $p$ the proper pressure, $\rho$ the proper total energy density,
and $U_\mu$ the velocity four-vector, defined so that $U_\mu U^\mu=-1$.
Since the fluid is at rest, we take $U^\mu=\frac{1}{\sqrt{B(r)}}\delta^\mu_t$.
We assume $p$ and $\rho$ are positive throughout the star.
Our assumptions of time independence and spherical symmetry imply
that $p$ and $\rho$ are functions only of the radial coordinate $r$.

Plugging Eq.~(\ref{eq:4a}) and Eq.~(\ref{eq:3a}) into Eq.~(\ref{eq:2a}), we
find that the equations read
\begin{eqnarray}
-G_t^t&=&\frac{D-2}{2}\frac{\tilde{\alpha}[(D-5)(A-1)^2A+2(A-1)A'r]
+(D-3)(A-1)A^2r^2+AA'r^3}{A^3 r^4}\nonumber\\
&=&\kappa\rho,\label{eq:5a}\\
G_r^r&=&\frac{D-2}{2}\frac{\tilde{\alpha}[-(D-5)(A-1)^2B+2(A-1)B'r]
-(D-3)(A-1)ABr^2+AB'r^3}{A^2Br^4}\nonumber\\
&=&\kappa p,\label{eq:6a}
\end{eqnarray}
where an prime denote the derivative with respect to $r$. Here we omit the third equation.
Instead, we use the equation for hydrostatic equilibrium which is
equivalent to the equation of energy-momentum conservation
\begin{eqnarray}
\frac{B'}{B}=-\frac{2p'}{p+\rho}.\label{eq:7a}
\end{eqnarray}
From Eq.~(\ref{eq:5a}) we find that the solution with $A(0)$ finite is
\begin{eqnarray}
A(r) = \frac{1}{1+\frac{r^2}{2\tilde{\alpha}}-\sqrt{\frac{r^4}{4\tilde{\alpha}^2}
+\frac{2\kappa M}{(D-2)\tilde{\alpha}\Omega_{D-2}r^{D-5}}}},\label{eq:8a}
\end{eqnarray}
where the mass function $M(r)$ is defined as $M(r)=\Omega_{D-2}\int_0^r{\rho(r')r'^{(D-2)}dr'}$.
Here we have abandoned another solution since it doesn't recover the
solution in general relativity in the limit $\alpha\to 0$.
The function $M(r)$ isn't the total energy of the matter since it
is defined as the integral of the energy density $\rho(r)$ of matter
alone thus doesn't include the energy of the gravitational field.
The thermal and gravitational energies of the star will be discussed in the next section.
Using Eq.~(\ref{eq:6a}), Eq.~(\ref{eq:7a}) and Eq.~(\ref{eq:8a}), we obtain
\begin{eqnarray}
p'=-r(p+\rho)\frac{\frac{2}{D-2}p+2\sqrt{\frac{1}{4\tilde{\alpha}^2}
+\frac{2\kappa M}{(D-2)\tilde{\alpha}\Omega_{D-2}r^{D-1}}}+\frac{2(D-5)\kappa M}{(D-2)\Omega_{D-2}r^{D-1}}
-\frac{1}{\tilde{\alpha}}}{4\tilde{\alpha } \sqrt{\frac{1}{4\tilde{\alpha}^2}
+\frac{2\kappa M}{(D-2)\tilde{\alpha}\Omega_{D-2}r^{D-1}}}(1
+\frac{r^2}{2\tilde{\alpha}}-\sqrt{\frac{r^4}{4\tilde{\alpha}^2}
+\frac{2\kappa M}{(D-2)\tilde{\alpha}\Omega_{D-2}r^{D-5}}})},\label{eq:9a}
\end{eqnarray}
and the solution with $B(\infty)=1$ is
\begin{eqnarray}
B(r')=\exp[-\int_{r'}^{+\infty}{r\frac{\frac{2}{D-2}p+2\sqrt{\frac{1}{4\tilde{\alpha}^2}
+\frac{2\kappa M}{(D-2)\tilde{\alpha}\Omega_{D-2}r^{D-1}}}+\frac{2(D-5)\kappa M}{(D-2)\Omega_{D-2}r^{D-1}}
-\frac{1}{\tilde{\alpha}}}{2\tilde{\alpha } \sqrt{\frac{1}{4\tilde{\alpha}^2}
+\frac{2\kappa M}{(D-2)\tilde{\alpha}\Omega_{D-2}r^{D-1}}}
(1+\frac{r^2}{2\tilde{\alpha}}-\sqrt{\frac{r^4}{4\tilde{\alpha}^2}
+\frac{2\kappa M}{(D-2)\tilde{\alpha}\Omega_{D-2}r^{D-5}}})}dr}].\label{eq:10a}
\end{eqnarray}

Outside the star, $p(r)$ and $\rho(r)$ vanish, and $M(r)$ is the constant $M(R)$,
therefore Eq.~(\ref{eq:8a}) and Eq.~(\ref{eq:10a}) give
\begin{eqnarray}
B(r)=A^{-1}(r)=1+\frac{r^2}{2\tilde{\alpha}}-\sqrt{\frac{r^4}{4\tilde{\alpha}^2}
+\frac{2\kappa M}{(D-2)\tilde{\alpha}\Omega_{D-2}r^{D-5}}}\label{eq:11a}
\end{eqnarray}
for arbitrarily $r\geq R$. The  Eq.~(\ref{eq:11a}) is the exterior solution of the static
spherically symmetric black hole in flat space. Hence, if matter can disperse to
the extra dimension, in the $D\geq5$ spherically symmetric space-time, there is
an interior solution which describes a perfect fluid relating to the exterior solution
of the static spherically symmetric black hole. Thus, Eq.~(\ref{eq:11a}) makes the
solution of black holes more believable since such exterior solution can be naturally
emerged by a suitable distribution of matter.

The Eq.~(\ref{eq:9a}) is the generalization of Tolman-Oppenheimer-Volkoff (TOV)
equation in the $D\geq5$ EGB theory. One can check that in the general relativity
limit $\alpha\to 0$, such an equation in 5D space-time will be written in the form of
\begin{eqnarray}
p'=-r(p+\rho)\frac{p+\frac{(D-3)\kappa M}{\Omega_{D-2}r^{D-1}}}{D-2-\frac{2\kappa M}{\Omega_{D-2}r^{D-3}}},\label{eq:12a}
\end{eqnarray}
which is similar to the TOV equation in 4D case \cite{Tolman:1949ne,Oppenheimer:1939ue,Oppenheimer:1939ne}. 
As in general relativity,
we can use this equation to obtain $\rho(r)$, $M(r)$, $p(r)$ throughout a  static spherically
symmetric star in the 5D space-time. The pressure $p$ may in general be expressed as a function
of the density $\rho$, the entropy per nucleon $s$, and the chemical composition. We can assume
that the entropy per nucleon $s$ does not vary throughout the star, and the star we consider have
a chemical composition that is constant throughout for simplicity, that is, $p(r)$ may be regarded
as a function of $\rho(r)$ alone. The definition of $M(r)$ provides an initial condition $M(0)=0$.
Eq.~(\ref{eq:9a}) together with an equation of state giving $p(\rho)$, serve to determine $\rho(r)$,
 $M(r)$, $p(r)$ of the star for arbitrarily $r$, once we specify the other initial condition that
 the value of $\rho(0)$. Eq.~(\ref{eq:9a}) should be integrated out from the center of the star,
 until $p(\rho(r))$ drops to zero at some point $r=R$, which we then interpret as the radius of the star.

\section{Effect of Gauss-Bonnet term\label{33s}}
\indent

This section is devoted to the influence on a spherically symmetric star
caused by Gauss-Bonnet term. First we investigate thermal and gravitational energies of the star.
As in general relativity, we can compare $M(R)$ with the energy $M_0$ that
the matter of the star would have  dispersed to infinity in $D\geq5$ space-time.
This is simply $M_0  = m_NN$, where $m_N$ is the rest-mass of a nucleon,
$N$ is the number of nucleons in the star which is given by
\begin{eqnarray}
N=\int_\Sigma{\sqrt{-g}J_N^0dr d\theta_1\cdots d\theta_{D-2}}=\Omega_{D-2}\int_0^R{\sqrt{A(r)B(r)}J_N^0r^{D-2}dr},\nonumber
\end{eqnarray}
where $J_N^\mu$ is the conserved nucleon current, $\theta_1,\cdots, \theta_{D-2}$ are spherical
coordinates on $S_{D-2}$, $\Sigma$ is a Cauchy surface of the spacetime. $J_N^0$ can be expressed
in terms of the proper nucleon number density $n$, which is $n=-U_\mu J_N^\mu=\sqrt BJ_N^0$.
The expression of $N$ then becomes
\begin{eqnarray}
N=\Omega_{D-2}\int_0^R{\sqrt{A(r)}n(r)r^{D-2}dr}.\label{eq:17a}
\end{eqnarray}
The proper number density $n(r)$ and $N$ are fixed for a star with a given constant
$s$ and chemical composition, once we choose $\rho(0)$. The internal energy of
 the  star is now defined by $E=M-m_NN$. We can define the proper internal material
 energy density as $e(r)=\rho(r)-m_Nn(r)$, and decompose the internal energy $E$ as the sum of
 the thermal and gravitational energies  $T$ and $V$, respectively
\begin{eqnarray}
T&=&\Omega_{D-2}\int_0^R{\sqrt{A}e(r)r^{D-2}dr},\nonumber\\
V&=&\Omega_{D-2}\int_0^R{(1-\sqrt{A})\rho(r)r^{D-2}dr}.\label{eq:22a}
\end{eqnarray}

Using Eq.~(\ref{eq:8a}), we obtain that
\begin{eqnarray}
T&=&\Omega_{D-2}\int_0^R{\frac{1}{\sqrt{1+\frac{r^2}{2\tilde{\alpha}}
-\sqrt{\frac{r^4}{4\tilde{\alpha}^2}+\frac{2\kappa M}{(D-2)\tilde{\alpha}
\Omega_{D-2}r^{D-5}}}}}e(r)r^{D-2}dr},\nonumber\\
V&=&\Omega_{D-2}\int_0^R{(1-\frac{1}{\sqrt{1+\frac{r^2}{2\tilde{\alpha}}
-\sqrt{\frac{r^4}{4\tilde{\alpha}^2}+\frac{2\kappa M}{(D-2)\tilde{\alpha}
\Omega_{D-2}r^{D-5}}}}})\rho(r)r^{D-2}dr}.\label{eq:24a}
\end{eqnarray}
It is straightforward to check that
\begin{eqnarray}
\frac{d A}{d\tilde{\alpha}}=-\frac{1}{2\tilde{\alpha}\sqrt{\frac{r^4}{4\tilde{\alpha}^2}+\frac{2\kappa M}
{(D-2)\tilde{\alpha}\Omega_{D-2}r^{D-5}}}}\Big(1-\frac{1}{1+\frac{r^2}{2\tilde{\alpha}}-\sqrt{\frac{r^4}
{4\tilde{\alpha}^2}+\frac{2\kappa M}{(D-2)\tilde{\alpha}\Omega_{D-2}r^{D-5}}}} \Big)^2<0,\label{eq:25a}
\end{eqnarray}
which leads to $\partial T/\partial \tilde{\alpha} <0$ and $\partial V/\partial \tilde{\alpha} >0$.
Hence, Gauss-Bonnet term with positive $\alpha$ will decreases the thermal
energy and increases the gravitational energy.

Next, we consider a stars with uniform density, consisting of incompressible fluids with $\rho=const$.
In this case Eq.~(\ref{eq:9a}) may be written as
\begin{eqnarray}
p'(r)=-r(p+\rho)\frac{\frac{2}{D-2}p+2\sqrt{\frac{1}{4\tilde{\alpha}^2}+\frac{2\kappa\rho}{(D-1)(D-2)\tilde{\alpha}}}
+\frac{2(D-5)\kappa\rho}{(D-1)(D-2)}-\frac{1}{\tilde{\alpha}}}
{4\tilde{\alpha}\sqrt{\frac{1}{4\tilde{\alpha}^2}+\frac{2\kappa\rho}{(D-1)(D-2)\tilde{\alpha}}}
[1+r^2(\frac{1}{2\tilde{\alpha}}-\sqrt{\frac{1}{4\tilde{\alpha}^2}
+\frac{2\kappa\rho}{(D-1)(D-2)\tilde{\alpha}}})]}.\label{eq:31a}
\end{eqnarray}
The solution with $p(R)=0$ is
\begin{eqnarray}
p(r)=\frac{\rho\Phi-\rho\Phi\Xi(r)}{\kappa\rho\Xi(r)-\Phi},\label{eq:32a}
\end{eqnarray}
here we define
\begin{eqnarray}
\Phi&=&(D-2)\sqrt{\frac{1}{4\tilde{\alpha}^2}+\frac{2\kappa\rho}{(D-1)(D-2)\tilde{\alpha}}}
+\frac{(D-5)\kappa\rho}{D-1}-\frac{D-2}{2\tilde{\alpha}},\nonumber\\
\Theta&=&\frac{1}{2\tilde{\alpha}}-\sqrt{\frac{1}{4\tilde{\alpha}^2}
+\frac{2\kappa\rho}{(D-1)(D-2)\tilde{\alpha}}},\nonumber\\
\Xi(r)&=&\sqrt{\frac{1+\Theta R^2}{1+\Theta r^2}}.\nonumber
\end{eqnarray}
Such solution doesn't make sense for all values of $\rho$ and $R$.
The pressure will become infinite at a point $r_\infty$ where
\begin{eqnarray}
r_\infty^2=\frac{(1+\Theta R^2)\kappa^2\rho^2-\Phi^2}{\Theta\Phi^2}.\nonumber
\end{eqnarray}
An infinity in $p(r)$ can't be blamed on an injudicious choice of coordinate system
since the pressure is a scalar. In order to ensure that $p(r)$ isn't singular
throughout the star, we find that
\begin{eqnarray}
1+\Theta R^2>\frac{\Phi^2}{\kappa^2\rho^2},\label{eq:33a}
\end{eqnarray}
which leads to $r_\infty^2<0$.

Hence, the function $B(R)=1+\Theta R^2$ has a lower limit.
This result has profound influence on the red shift of spectral lines from the surface of stars.
The red shift $z$ will have an upper bound $z_{max}$
since $z=\frac{\vartriangle\lambda}{\lambda}=B^{-\frac{1}{2}}(R)-1$, which gives
\begin{eqnarray}
z_{max}=\frac{\kappa\rho}{\Phi}-1.\label{eq:33c}
\end{eqnarray}
In the general relativistic limit, we have $\mathop {\lim }\limits_{\alpha \to 0}z_{max}=2/(D-3)$.
Obviously, non-negative
coefficient $\alpha$ will change the upper bound of the red shift, which is given under  in FIG.1.
The most interesting consequence of the of the second order
curvature corrections is that $z_{max}$ is dependent on the value of $\rho$
while it is a constant in general relativity case \cite{Bondi:1964zza}.
\begin{figure}
\centering \subfigure[$D=5$]{
\label{fig:subfig:A} 
\includegraphics{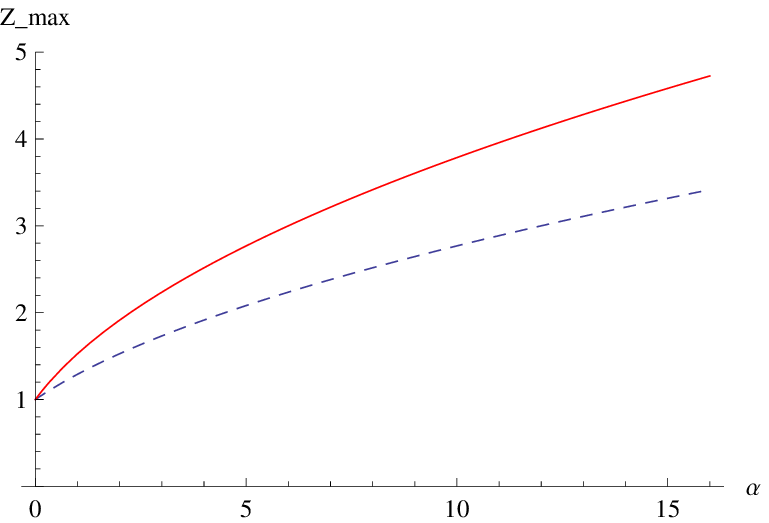}}%
\hfill%
\subfigure[$D=6$]{
\label{fig:subfig:B} 
\includegraphics{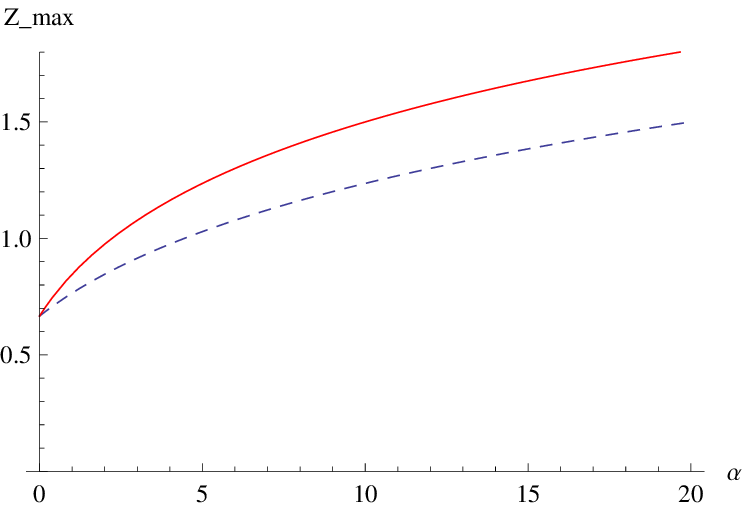}}
\caption{ Relation between $\alpha$ and the upper bound of $z_{max}$ in 5D and 6D Gauss-Bonnet gravity:
the solid and dotted lines representing $\rho=2$ and  $\rho=1$ respectively.
In general relativity limit $\alpha\to 0$ two lines have a same limit.. }
\label{fig:subfig} 
\end{figure}

\section{Stability of equilibrium\label{44s}}
\indent

Our solution represents an equilibrium state of the star, but it may be stable or unstable.
For most purposes, we only concerned with the stable solution. In order
to investigate the stability of a particular configuration, it would be necessary to compute the
frequencies $\omega_n$ of all normal modes of the configuration. If the frequency
$\omega_n$ has a positive imaginary part, the factor $exp(-i\omega_nt)$
would grow exponentially, and the system would be unstable. However, it is often possible
to obtain from the equilibrium solution alone whether the corresponding
configuration is stable, since we have following two theorems:

1. A star consisting of a perfect fluid with constant chemical composition and entropy per
nucleon can only pass from stability to instability with respect to some particular radial
normal mode, at a value of the central density $\rho(0)$ for
which the equilibrium energy $E$ and nucleon number $N$ are stationary that
\begin{eqnarray}
\frac{\partial E(\rho(0),s\cdot\cdot\cdot)}{\partial\rho(0)}&=&0,\nonumber\\
\frac{\partial N(\rho(0),s\cdot\cdot\cdot)}{\partial\rho(0)}&=&0,\label{eq:26a}
\end{eqnarray}
by a radial normal mode is meant a mode of oscillation in which the density
perturbation $\delta\rho$ is a function of $t$ and $r$ alone, and in which
nuclear reactions, viscosity, heat conduction, and radiative energy transfer play no role.

2. A particular star configuration with uniform entropy per nucleon and
chemical composition will satisfy Eq.~(\ref{eq:9a}) for equilibrium,
if and only if the quantity $M$ defined by $M=\Omega_{D-2}\int{\rho(r)r^3dr}$
is stationary with respect to all variations of $\rho(r)$ that leave
unchanged the quantity $N=\Omega_{D-2}\int{\sqrt An(r)r^3dr}$ and that
leave the entropy per nucleon and the chemical composition uniform and unchanged.
The equilibrium is stable with respect to radial oscillations if and only if $M$
or equivalently $E$ is a minimum with respect to all such variations.

These theorems have been proven in general relativity in
4D space-time \cite{Harrison:1984rg,Morse:1953rg}.
The proof of theorems $1$ is unconcerned with dimensions and the metric of space-time,
therefore it is obviously
appropriate for our $D\geq5$ star in Gauss-Bonnet gravity.
Now we are going to prove theorems $2$ in present case by using Lagrange
multiplier method. $M$ will be stationary with respect
to all variations that leave $N$ fixed if and only if there
exists a constant $\lambda$ for which $M-\lambda N$
is stationary with respect to all variations. In general,
the change in $M-\lambda N$ for a given variation $\delta\rho(r)$ is
\begin{eqnarray}
\delta M-\lambda\delta N&=&\Omega_{D-2}\int_0^\infty{\delta \rho(r)r^{D-2}dr}
-\lambda\Omega_{D-2}\int_0^\infty{\frac{\delta n(r)r^{D-2}dr}{\sqrt{1
+\frac{r^2}{2\tilde{\alpha}}-\Sigma(r,M)}}}\nonumber\\
&-&\lambda\Omega_{D-2}\int_0^\infty{\frac{n(r)\delta M(r)r^3dr}{2(D
-2)\tilde{\alpha}\Omega_{D-2}\Sigma(r,M)(1+\frac{r^2}{2\tilde{\alpha}}
-\Sigma(r,M))^{\frac{3}{2}}}},\label{eq:27a}
\end{eqnarray}
where $\Sigma(r,M)=\sqrt{\frac{r^4}{4\tilde{\alpha}^2}+\frac{2\kappa M}{(D-2)\tilde{\alpha}
\Omega_{D-2}r^{D-5}}}$. Here the integrals are carried to infinity for notational
convenience since they vanish outside a radius $R+\delta R$.
These variations are supposed not to change the entropy per nucleon,
that is $0=\delta(\frac{\rho}{n})+p\delta(\frac{1}{n})$,
therefore $\delta n(r)=\frac{n(r)}{p(r)+\rho(r)}\delta\rho(r)$.
Obviously, $\delta M(r)=\Omega_{D-2}\int_0^r{\delta\rho(r')r'^{D-2}dr'}$.
Thus, $\delta M-\lambda\delta N$ will vanish for all $\delta\rho(r)$ if and only if
\begin{eqnarray}
\frac{1}{\lambda}&=&\frac{n}{(p+\rho)\sqrt{1+\frac{r^2}{2\tilde{\alpha}}-\Sigma(r,M)}}\nonumber\\
&+&\Omega_{D-2}\int_r^\infty{\frac{n(r')r'^3dr'}{2(D
-2)\tilde{\alpha}\Omega_{D-2}\Sigma(r',M)(1+\frac{r'^2}{2\tilde{\alpha}}
-\Sigma(r',M))^{\frac{3}{2}}}}.\label{eq:29a}
\end{eqnarray}
This will be the case for some Lagrange multiplier $\lambda$ if and only
if the right-hand side is independent of $r$ that
\begin{eqnarray}
0&=&[\frac{n'}{p+\rho}-\frac{n(p'+\rho')}{(p+\rho)^2}]\frac{1}{\sqrt{1
+\frac{r^2}{2\tilde{\alpha}}-\Sigma(r,M)}}\nonumber\\
&+&\frac{n}{p+\rho}\frac{\frac{r^3}{2\tilde{\alpha}}+\frac{\rho r^3}{D-2}
-\frac{(D-5)\kappa M}{(D-2)\Omega_{D-2}r^{D-4}}
-r\Sigma(r,M)}{2\tilde{\alpha}\Sigma(r,M)(1+\frac{r^2}{2\tilde{\alpha}}
-\Sigma(r,M))^{\frac{3}{2}}}\nonumber\\
&-&\frac{nr^ 3}{2(D-2)\tilde{\alpha}\Sigma(r,M)(1+\frac{r^2}{2\tilde{\alpha}}
-\Sigma(r,M))^{\frac{3}{2}}}.\label{eq:30a}
\end{eqnarray}
The condition of uniform entropy per nucleon gives $0=(\frac{\rho}{n})'+p(\frac{1}{n})'$
and therefore $n'(r)=\frac{n(r)\rho'(r)}{p(r)+\rho(r)}$. Substituting the expression of
$n'(r)$ in to Eq.~(\ref{eq:30a}), we obtain Eq.~(\ref{eq:9a}). Consequently $\delta M$ vanishes
for all $\delta\rho(r)$ under a given $ N$  and  Eq.~(\ref{eq:9a}). Thus, the stability
is  determined by the sign of second order of  $\delta M$. Namely, the star is stable if
 $\delta^2M(r)/\delta\rho(r)^2\geq 0$  and unstable if $\delta^2M(r)/\delta\rho(r)^2\leq 0$.

\section{Closing remarks\label{55s}}
\label{33s}

In this paper, we have investigated static spherically symmetric
stars in the $D\geq5$ Gauss-Bonnet gravity without the cosmological constant.
It is shown that there is an exact solution of such system
relating to the exterior solution of the static spherically symmetric black hole,
and the Gauss-Bonnet interaction decreases or enhances the thermal energy
and increases the gravitational energy of stars respectively.
The Gauss-Bonnet terms will modify the upper bound of red shift of spectral lines
from the surface of stars of uniform density, and the upper bound is dependent
on the value of density rather than  a constant in general relativity case.
Besides, two theorems for judging the stability of equilibrium of
stars in $D\geq5$ Gauss-Bonnet gravity are proved, which
are the natural generalization of ones  in general relativity.

\end{document}